# REACTION-DIFFUSION MODELS FOR GLIOMA TUMOR GROWTH

Miguel Martín-Landrove*

Center for Medical Visualization, National Institute for Bioengineering (INABIO), Universidad Central de Venezuela.

mglmrtn@yahoo.com

* Member of the Physics & Mathematics in Biomedicine Consortium

**Abstract.**

Mathematical modelling of tumor growth is one of the most useful and inexpensive approaches to determine and predict the stage, size and progression of tumors in realistic geometries. Moreover, these models has been used to get an insight into cancer growth and invasion and in the analysis of tumor size and geometry for applications in cancer treatment and surgical planning. The present revision attempts to present a general perspective of the use of models based on reaction-diffusion equations not only for the description of tumor growth in gliomas, addressing for processes such as tumor heterogeneity, hypoxia, dormancy and necrosis, but also its potential use as a tool in designing optimized and patient specific therapies.

**Resumen.**

El modelaje matemático del crecimiento tumoral representa uno de los métodos más útiles y de más bajo costo para determinar y predecir tanto el estadio, tamaño y evolución de los tumores en geometrías reales. Más aún, estos modelos han sido usados para comprender el crecimiento e invasión del cáncer y en el análisis del tamaño del tumor y su geometría para su aplicación en el tratamiento del cáncer y la planificación quirúrgica. La presente revisión intenta presentar desde una perspectiva general el uso de modelos basados en ecuaciones de reacción difusión no solamente para la descripción del crecimiento tumoral en gliomas, con énfasis en procesos tales como heterogeneidad tumoral, hipoxia, latencia o necrosis, sino también su uso potencial en el diseño optimizado de terapias específicas al paciente.

**Palabras Clave.**

Ecuación Reacción Difusión, Crecimiento Tumoral, Optimización, Terapia del Cáncer.

**Introduction.**

Gliomas are primary brain tumors that arise from precursors of glial cells in the brain. Among them, glioblastomas are the most aggressive and as it happens with most of the gliomas, its growth is so fast and extensive that patients do not exhibit any symptoms imposing serious limitations for treatment. Depending on its grade, gliomas exhibit internal necrotic regions and a periphery of active tumor cells that invades the surrounding tissue. In the case of glioblastomas or high grade gliomas, images obtained by contrast enhanced CT or MRI commonly resembles the tumor growth as a kind of explosion, i.e., a fast growing front wave of very active tumor cells that expands and leaves behind hypoxic, hypoglycemic or necrotic tissues. This type of growth can be described by reaction-diffusion equations or ballistic growth models, or proliferative-invasive models in the biological context. In the following sections several reaction-diffusion models will be discussed, how the effect of oxygenation and nutrient concentration has been addressed to account for cell hypoxia, tumor dormancy and necrosis, how model parameters are extracted from different medical imaging modalities and how all this knowledge has led to the proposal of working models to simulate and optimize different therapies.

**Reaction-Diffusion Basic Model.**

The first attempts to model glioma tumor growth by means of a reaction-diffusion mathematical model [1] were performed by Cruywagen et al. [2], Tracqui et al. [3] and Woodward et al. [4] in order to account for the effect of therapies on glioma growth and later by Burgess et al. [5] to emphasize the importance of diffusion on glioma growth. The model is described by a partial differential equation [1],

$$\frac{\partial c}{\partial t} = -\nabla \cdot \vec{J} + S(c,t) - T(c,t) \qquad (1)$$

where $c$ is the tumor cell concentration, $\vec{J}$ is the tumor cellular flux, $S(c,t)$ is a term that accounts for cellular proliferation and $T(c,t)$ represents the contribution of treatment. Assuming that the cellular flux obeys Fick's law,

$$\vec{J} = -D\nabla c \qquad (2)$$

and the proliferation term is set to produce an exponential growth, then equation (1) leads to

$$\frac{\partial c}{\partial t} = D\nabla^2 c + \rho c \qquad (3)$$

where $D$ is a diffusion coefficient that accounts for tumor invasiveness and $\rho$ is the tumor cell proliferation rate. The solution of equation (3) is restricted by the boundary condition that the flux of cells outside the brain or into the ventricles is zero, that is,

$$\vec{n} \cdot \nabla c = 0 \qquad (4)$$

where $\vec{n}$ is a unitary vector normal to the cortical and ventricular surfaces. Experiments performed on rats demonstrated that glioma cells disperse more effectively along white matter axon tracts [6-8] than along neuronal cell bodies in gray matter, which leads to a variation of equation (3), proposed by Swanson et al. [9,10], which includes the spatial dependence of the diffusion coefficient $D$,

$$\frac{\partial c}{\partial t} = \nabla \cdot (D\nabla c) + \rho c \qquad (5)$$

To evaluate the differences between grey and white matter motilities, Swanson et al. [9,10] used the Fisher approximation [1,11] which stablishes that a travelling wave solution of equation (3) propagates with a terminal velocity given by,

$$v = 2(\rho D)^{1/2} \qquad (6)$$

Equation (6) allows for the estimation of the diffusion coefficient knowing the wave front propagation velocity and the proliferation rate, $\rho$. Fisher's approximation has been used extensively for parameter extraction of gliomas by Mandonnet et al. [12]. Wave front velocities were obtained from the analysis of CT scans [2-5,12] and finally, assuming a proliferation rate $\rho = 0.012 \ day^{-1}$, the diffusion coefficients can be estimated [9,10] as $D_g = 0.0013 \ cm^2/day$, and $D_w = 5D_g = 0.0065 \ cm^2/day$. Simulations of the glioma growth were performed by Swanson et al. [9-10,13-15] on a virtual space using the Brainweb database [16,17], assuming a threshold concentration of $8000 \ cells/mm^3$ [5], The results of these simulations are shown in Figure 1, where it can be appreciated that tumor extension greatly exceeds what is detected by contrast enhanced CT [5], depicted as thick black contours. Equation (5) predicts an exponential growth of tumor cell concentration, which at longer times achieves a value that is no longer realistic or sustainable at cellular level due to the scarcity of oxygen and nutrients. This problem can be easily solved by assuming that there is a limit concentration, $c_m$, and the proliferative term in equation (5) is changed by a logistic or Verhulst growth model,

$$\frac{\partial c}{\partial t} = \nabla \cdot (D\nabla c) + \rho c \left(1 - \frac{c}{c_m}\right) \qquad (7)$$

From now on, this model is called the Proliferative Invasive model or PI model and represents the starting point for the proposal of more realistic models.

**Adding more reality to the model.**

One of the advantages of equation (7) is that it is parametrized with few parameters which can be easily extracted from patient data, being the main difficulty the computational implementation to solve it on reliable virtual environments. Some additional considerations have to be done in order to obtain a glioma growth model that takes into account the anisotropy of the diffusion tensor, which is particularly relevant in brain tissue, the effect of mechanical deformations or mass effect and tumor heterogeneity.

   a) **Anisotropy of the Diffusion Tensor.**

The diffusive term in equation (7) not only exhibits a spatial variation depending if brain tissue corresponds to either gray or white matter, but it has tensor properties as well, so gray matter is mostly isotropic while white matter is highly anisotropic, In the case of white matter, diffusion along the fiber axon tracts is $1.2 \times 10^{-3}\ mm^2 s^{-1}$ while it is $0.4 \times 10^{-3}\ mm^2 s^{-1}$ perpendicularly [18]. Taking this fact into account, equation (7) can be rewritten as [19-21],

$$\frac{\partial c}{\partial t} = \sum_i \sum_j D_{ij} \frac{\partial^2 c}{\partial X_i \partial X_j} + \sum_i \sum_j \frac{\partial D_{ij}}{\partial X_i} \frac{\partial c}{\partial X_j} + \rho c \left(1 - \frac{c}{c_m}\right)$$

$$(8)$$

for highly anisotropic white matter, and in the case of gray matter which is nearly isotropic, i.e., $D_{ij} = D\delta_{ij}$,

$$\frac{\partial c}{\partial t} = D\nabla^2 c + \nabla D \cdot \nabla c + \rho c \left(1 - \frac{c}{c_m}\right) \quad (9)$$

which corresponds to the model described by equation (7) where only the spatial heterogeneity [9,10] is considered. Simulations were performed by Jbabdi et al. [21] using equation (8) and diffusion tensor MRI (DT-MRI) data [22-24]. Since DT-MRI only maps the diffusion of water molecules, and tumor cell diffusion along white matter axon tracts is likely to be more anisotropic [25,26], some assumptions were made to determine tumor cell diffusion in equation (8) starting from DT-MRI data, for instance, a scaling factor that preserves the axon tract direction but changes the diffusion tensor magnitude and anisotropy. The diffusion tensor can be transformed to its diagonal form [27] and can be written as,

$$\widetilde{D}_w = \lambda_1 \vec{v}_1 \vec{v}_1^T + \lambda_2 \vec{v}_2 \vec{v}_2^T + \lambda_3 \vec{v}_3 \vec{v}_3^T \quad (10)$$

where $\lambda_i$ and $\vec{v}_i$ are the eigenvalues and eigenvectors, respectively, of the diffusion tensor in the principal axis reference frame. The tensor can also be characterized by three indices [27], $c_l$, $c_p$ and $c_s$, defined in terms of the eigenvalues $\lambda_i$, assuming that $\lambda_1 \geq \lambda_2 \geq \lambda_3$,

$$c_l = \frac{\lambda_1 - \lambda_2}{\lambda_1 + \lambda_2 + \lambda_3}, c_p = \frac{2(\lambda_2 - \lambda_3)}{\lambda_1 + \lambda_2 + \lambda_3}, c_s = \frac{3\lambda_3}{\lambda_1 + \lambda_2 + \lambda_3}$$

(11)

with $c_l + c_p + c_s = 1$. Depending on the eigenvalues $\lambda_i$, the tensor is linear, $c_l = 1$, planar, $c_p = 1$, or spherical $c_s = 1$, the isotropic case, i.e., gray matter. The scaled tensor can be defined as [21],

$$\widetilde{D} = a_1(r)\lambda_1 \vec{v}_1 \vec{v}_1^T + a_2(r)\lambda_2 \vec{v}_2 \vec{v}_2^T + a_3(r)\lambda_3 \vec{v}_3 \vec{v}_3^T$$

(12)

where,

$$\begin{bmatrix} a_1 \\ a_2 \\ a_3 \end{bmatrix} = \begin{bmatrix} r & r & 1 \\ 1 & r & 1 \\ 1 & 1 & 1 \end{bmatrix} \begin{bmatrix} c_l \\ c_p \\ c_s \end{bmatrix} \quad (13)$$

For the simulations, $r = 10$ was used [21]. A more general approach, assuming a proportionality between tumor cell diffusion anisotropy and diffusion tensor fractional anisotropy was proposed by Painter et al. [28],

$$k = \kappa\, FA(\widetilde{D}) \quad (14)$$

With, $FA(\widetilde{D})$, the fractional anisotropy of the diffusion tensor [22-24,27], defined as,

$$FA(\widetilde{D}) = \frac{\sqrt{(\lambda_1 - \lambda_2)^2 + (\lambda_2 - \lambda_3)^2 + (\lambda_1 - \lambda_3)^2}}{\sqrt{2(\lambda_1^2 + \lambda_2^2 + \lambda_2^2)}}$$

(15)

where $\lambda_i$ are the diffusion tensor eigenvalues Results of simulations carried on real Diffusion Tensor Imaging or DTI are shown in Figure 2, showing a strong dependence on the anisotropy enhancement parameter, $\kappa$. To evaluate the correspondence of actual tumor DT images with the simulations, first Moyasebi et al. [29] and later, Swan et al. [30], used the Jaccard index which is a measurement of similarity between two finite data sets, i.e. simulated tumor growth and actual tumor DTI, and defined as,

$$J(ST, TDTI) = \frac{|ST \cap TDTI|}{|ST \cup TDTI|} \quad (16)$$

with, $ST$ and $TDTI$, the simulated tumor and the actual tumor DTI, respectively. The Jaccard index was used to optimize tumor growth parameters such as the initial starting point, proliferation rates and anisotropy enhancement parameter [30]. The initial starting point was determined by computing the Jaccard index for simulations starting at different points and selecting the one with the maximum value of the index. Similarly, the tumor growth time was determined by computing the Jaccard index along the time evolution in a simulation, as shown in Figure 3.

**b) Biomechanical Deformations.**

The first attempt to include mechanical deformations to the glioma growth model were introduced by Clatz et al [19,20,31]. They used rheological brain properties [32-34] to derive the brain constitutive equation. Tumor growth occurs very slowly, so a static equilibrium equation can be proposed for the biomechanical deformations,

$$\nabla \cdot \tilde{\sigma} + \vec{f}_{ext} = 0 \qquad (17)$$

where $\tilde{\sigma}$ is the internal stress tensor and $\vec{f}_{ext}$ is the external force. Since the growing process is very slow, it can also be assumed that there are linear relationships for the constitutive equation and the strain computation,

$$\tilde{\sigma} = \widetilde{K}\tilde{\epsilon} \qquad (18)$$

$$\tilde{\epsilon} = \frac{1}{2}\left(\widetilde{\nabla \vec{u}} + \widetilde{\nabla \vec{u}}^T\right) \qquad (19)$$

where $\widetilde{K}$ is the elasticity tensor, $\tilde{\epsilon}$ is the linearized Lagrange strain tensor and $\vec{u}$ is the tissue displacement. Following Wasserman et al. [35], they proposed a modified equilibrium equation to take into account the mechanical impact of the tumor growth (mass effect) on the surrounding tissue, with a term proportional to the tumor concentration, $c$,

$$\nabla \cdot (\tilde{\sigma} - \alpha c \tilde{I}) + \vec{f}_{ext} = 0 \qquad (20)$$

Some of the results are shown in Figure 4, where it can be appreciated that the higher the tumor cell concentration is, shown in Figure 4a, the stronger is the mass effect, which corresponds to high tissue displacements or deformations, as shown in Figure 4b. More generally, Hogea et al. [36-38] proposed a general mass balance equation for the tumor growth,

$$\frac{\partial c}{\partial t} = \nabla \cdot (\widetilde{D}\nabla c) - \nabla \cdot (c\vec{v}) + \rho c(1 - c) \qquad (21)$$

where the tumor cell concentration has been normalized to the maximum $c_m$ and an advection term has been added which includes a drift velocity, $\vec{v}$ which depends on tumor specific mechanisms such as, for example, chemotaxis [36]. The stress tensor is defined in a similar way to references [19,20,31], as follows,

$$\tilde{\sigma} = (\lambda \nabla \cdot \vec{u})\tilde{I} + \mu(\widetilde{\nabla \vec{u}} + \widetilde{\nabla \vec{u}^T}) \qquad (22)$$

where $\lambda$ and $\mu$ are Lame`s coefficients, related to Young`s modulus $E$ and Poisson's ratio $v$. The equilibrium equation (20) is modified to,

$$\nabla \cdot \left((\lambda \nabla \cdot \vec{u})\tilde{I} + \mu(\widetilde{\nabla \vec{u}} + \widetilde{\nabla \vec{u}^T})\right) - f(c, \vec{p})\nabla c = 0 \qquad (23)$$

where the proportionality factor, $f(c, \vec{p})$, is positive and as proposed in references [36-38], can be parametrized as,

$$f(c, \vec{p}) = p_1 e^{-\frac{p_2}{c^s}} e^{-\frac{p_2}{(2-c)^s}} \qquad (24)$$

with $\vec{p} = (p_1, p_2, s)$. Equation (24) is a monotonically increasing function in the concentration range $0 < c \leq 1$ with its maximum at $c = 1$. To complement equations (21) and (23), the following set of equations have to be added,

$$\vec{v} = \frac{\partial \vec{u}}{\partial t} \qquad (25)$$

$$\frac{\partial \vec{m}}{\partial t} + (\vec{v} \cdot \nabla)\vec{m} = 0 \qquad (26)$$

where $\vec{m}$ stands for $(\lambda, \mu, \widetilde{D})$. Equations (21), (23), (25) and (26) were used by Gooya et al. [39] to perform atlas registration of simulated gliomas with patient images, in order to evaluate tumor location, mass effect and degree of infiltration, with results shown in Figure 5.

### c) Tumor Heterogeneity, Hypoxia, Necrosis and Angiogenesis.

The World Health Organization (WHO) grading scheme for gliomas takes into account variations in tumor cellularity, mitoses and vascular proliferation. In particular, the characteristic vascularity of high grade gliomas, such as glioblastomas, and the relation to glioma growth to neo vascularity or angiogenesis is a key feature in the modeling of glioma growth [40-42]. Swanson et al. [43] proposed a model that included angiogenesis together with hypoxia and necrosis in order to quantify the role of angiogenesis in the malignant progression of gliomas. The starting equation is given by equation (7) but with certain modifications,

$$\frac{\partial c}{\partial t} = \nabla \cdot (D(1-T)\nabla c) + \rho c(1-T) + \gamma hV$$
$$-\beta c(1-V) - \alpha_n nc \qquad (27)$$

where,

$$T = \frac{(c+h+v+n)}{c_m} \qquad (28)$$

$$V = \frac{v}{(c+h+v)} \qquad (29)$$

and $h$, $v$ and $n$ are the hypoxic, vascular and necrotic cell concentrations, respectively. The first two terms in equation (27) account for the dispersion and proliferation of normoxic glioma cells, the third corresponds to the conversion of hypoxic to normoxic glioma cells, the fourth to the conversion of normoxic to hypoxic glioma cells and the fifth to the conversion of normoxic to necrotic glioma cells, that depends on necrotic cell concentration through a factor $\alpha_n$ [44]. Similar equations can be written for $h$, $n$ and $v$, the hypoxic, necrotic and vascular, respectively,

$$\frac{\partial h}{\partial t} = \nabla \cdot (D(1-T)\nabla h) - \gamma hV + \rho c(1-V)$$
$$-(\alpha_h h(1-V) + \alpha_n nh) \qquad (30)$$

with the first term corresponding to hypoxic glioma cell dispersion, the second to the conversion of hypoxic to normoxic glioma cells, the third to the conversion of normoxic to hypoxic glioma cells and the fourth, to the conversion of hypoxic to necrotic glioma cells,

$$\frac{\partial n}{\partial t} = \alpha_h h(1-V) + \alpha_n n(c+h+v) \quad (31)$$

$$\frac{\partial v}{\partial t} = \nabla \cdot (D_v(1-T)\nabla v) + \mu \frac{a}{K_m+a} v(1-T) - \alpha_n nv \quad (32)$$

where the first term is the dispersal of vasculature, i.e. endothelial cells, characterized by a diffusion coefficient $D_v$ [45], the second corresponds to the vasculature proliferation that depends on the concentration of angiogenic factors, $a$, with a maximal proliferation rate $\mu$ [46], and the third term corresponds to the conversion of vasculature to necrotic tissue. Finally the equation for the angiogenic factors concentration is,

$$\frac{\partial a}{\partial t} = \nabla \cdot (D_a \nabla a) + \delta_c c + \delta_h h - \lambda a$$

$$- q\mu \frac{a}{K_m+a} v(1-T) - \overline{\omega} av \quad (33)$$

Which includes the diffusion of the angiogenic factors with $D_a$ [47], the production of these factors by normoxic cells [48], at a rate $\delta_c$ and by hypoxic cells (VGEF), at a rate $\delta_h$ [49] and its decay with rate $\lambda$ [50]. The last two terms in equation (33) correspond to the net consumption of angiogenic factors by the vasculature. Main results are shown in Figure 6. Notice that as tumor grading increases there is an enhancement of normoxic and hypoxic tumor cell concentrations, as well as endothelial cell and VGEF concentrations close to the tumor interface, particularly for Grade IV gliomas, Figure 6a.

Some variations to the previous scheme were proposed by Papadogiorgaki et al. [51] by assuming a multi-compartmental model of coupled reaction diffusion equations, each one related to compartments that are spatially distributed along the tumor from its periphery to its inner necrotic core as shown in Figure 7. Each compartment is characterized by glioma cell viability and phenotype and, starting at the tumor outer interface, there is a proliferative compartment, followed by a hypoxic, hypoglycemic and necrotic compartments, all of them embedded in the extracellular matrix or ECM, which supplies for oxygen and glucose to the tumor compartments and is destroyed by matrix-degradative enzymes or MDEs [52] which are secreted by proliferative and hypoxic glioma cells. Increased glucose consumption by cancer cells, known as the Warburg effect [53], with independence on oxygen levels justifies the inclusion of the hypoglycemic compartment in a separate way to the hypoxic compartment [52-55]. The diffusion equations for each compartment are,

$$\frac{\partial c}{\partial t} = \nabla \cdot (D_c(1-T)\nabla c) + \tilde{\rho}c(1-T) + g_h(1-n_h)h$$
$$+ g_q(1-gl_q)q - b_h n_h c - b_q gl_q c - a_n nc$$

(34)

where $\tilde{\rho}$ is a modified proliferation rate that depends on oxygen and glucose concentrations [51], $g_h$, $g_q$ are the conversion rates from hypoxic and hypoglycemic to proliferative, respectively, $b_h$, $b_q$, the conversion rates from proliferative to hypoxic or hypoglycemic, respectively, and $n_h$, $gl_q$ are variables that depend on the oxygen and glucose concentration thresholds, such that they take the value 0 if the concentration is above the threshold and 1 otherwise, finally, as in equation (27), there is a term to account for conversion to the necrotic compartment and $T$ is similarly defined as in equation (28),

$$T = \frac{(c+h+q+n)}{c_m}$$

(35)

Similarly, for the hypoxic compartment,

$$\frac{\partial h}{\partial t} = \nabla \cdot (D_h(1-T)\nabla h) + b_h n_h c - g_h(1-n_h)h - a_h n_{hn} h$$
$$- a_{glh} gl_{hn} h - a_n nh$$

(36)

The last three terms in equation (36) correspond to conversion to the necrotic compartment, depending on oxygen and glucose threshold concentrations [51] though the coefficients $n_{hn}$ and $gl_{hn}$, as,

$$n_{hn} = \begin{cases} 0, & \alpha_h \leq 0.9 \\ 1, & \alpha_h > 0.9 \end{cases} \qquad gl_{hn} = \begin{cases} 0, & n_h = 1, \ gl_q = 0 \\ 1, & n_h = 1, \ gl_q = 1 \end{cases}$$

(37)

where,

$$\alpha_h = \frac{h}{c+h+q} \tag{38}$$

For the hypoglycemic compartment,

$$\frac{\partial q}{\partial t} = \nabla \cdot (D_q(1-T)\nabla q) + b_q gl_q c - g_q(1-gl_q)q - a_q n_{qn} q - a_{glq} gl_{qn} q - a_n nq$$

(39)

and similarly [51],

$$gl_{qn} = \begin{cases} 0, & \alpha_q \leq 0.75 \\ 1, & \alpha_q > 0.75 \end{cases} \qquad n_{qn} = \begin{cases} 0, & n_h = 0, \ gl_q = 1 \\ 1, & n_h = 1, \ gl_q = 1 \end{cases}$$

(40)

with

$$\alpha_h = \frac{q}{c+h+q} \tag{41}$$

Finally, for the necrotic compartment,

$$\frac{\partial n}{\partial t} = a_n nc + a_h n_{hn} h + a_{glh} gl_{hn} h + a_n nh + a_q n_{qn} q + a_{glq} gl_{qn} q + a_n nq$$

(42)

The remaining equations are for the oxygen, glucose, ECM and MDE concentrations, respectively,

$$\frac{\partial o}{\partial t} = D_o \nabla^2 o + \beta_o e - \alpha_o o - \gamma_{co} c - \gamma_{ho} h - \gamma_{qo} q \tag{43}$$

$$\frac{\partial g}{\partial t} = D_g \nabla^2 g + \beta_g e - \alpha_g g - \gamma_{cg} c - \gamma_{hg} h - \gamma_{qg} q \tag{44}$$

$$\frac{\partial e}{\partial t} = -\delta m e \tag{45}$$

$$\frac{\partial m}{\partial t} = D_m \nabla^2 m + \mu_c c + \mu_h h + \mu_q q - \lambda m \tag{46}$$

As it is readily seen, models given by [43,51] are extremely detailed and many parameters are needed to describe the entire model. A simpler approach can be proposed [58,59] where only a reaction diffusion equation is considered, i.e., the one associated to tumor normoxic cells, and those compartments related to hypoxic, hypoglycemic and necrotic cells are extracted from the original equation by certain rules [58,59], depending on tumor cell concentration and glucose levels. As the tumor interface progresses, diffusion of nutrients is less effective to account for all the cellular energetic requirements, particularly if it attempts to remain in a proliferative-invasive state, and as a consequence, for a certain nutrient concentration threshold value a transition to a hypoglycemic state occurs, and the cell no longer participates in the proliferation-invasion equation for normoxic tumor cells. This condition can occur when cell concentrations in a surrounding neighborhood around a particular volume element or voxel attain almost saturation values $c_m$, therefore a scarcity of resources is present, and there is some effective distance to tumor interface that is related to the diffusion length for nutrients (glucose), $\delta_G$, that can be estimated from the glucose diffusion coefficient $D_G$ of $5.79\ mm^2/day$ [58]. To evaluate the glucose concentration the following parameter is introduced [58,59],

$$\langle f_j \rangle = \frac{1}{N_{V_R}} \sum_{i \in V_R} (1 - c_i) \times e^{-\frac{d_{ij}^2}{\delta_G^2}} \tag{47}$$

where $V_R$ is a spherical volume of radius $R > \delta_G$ around the voxel $j$, $d_{ij}$ is the distance between voxels and $N_{V_R}$ is the total number of voxels inside the volume $V_R$, as defined in Figure 8. The state of each voxel in a three-dimensional simulation depends on the values of its concentration $c_i$ and the parameter $\langle f_i \rangle$, according to Table 1, which shows the transition rules between the different compartments. Results of this model are shown in Figure 8. Moreover, the model can be further simplified, Patel et al. [61] proposed a simple model, suitable to describe high grade gliomas, such as glioblastoma multiforme, that assume only two compartments, the proliferative-invasive or normoxic compartment and the necrotic compartment,

$$\frac{\partial c}{\partial t} = \begin{cases} \nabla \cdot (\widetilde{D} \nabla c) + \rho c(1-c) & max(c) < \tau \\ -\eta c & max(c) \geq \tau \end{cases} \quad (48)$$

where $\tau$ is a threshold for the cellular concentration and $\eta$ is the rate at which tumor cell concentration decay exponentially after this threshold is exceeded. Both quantities $\tau, \eta \in (0,1)$ and for the simulations the following values were used $\tau = 0.85, \eta = 0.90$ for best results fitted with patient data [61]. Part of the results are shown in Figure 10.

**Simulation of Therapy.**

Treatment of gliomas is commonly initiated by tumor resection, followed by radiotherapy and chemotherapy either separately or in combination. The reaction-diffusion model was used for tumor resection therapy [4,62,63], chemotherapy [3,64,65] and radiation therapy [59,66-75]. Commonly, the effect of therapy is included in the differential equation as,

$$\frac{\partial c}{\partial t} = \nabla \cdot (\widetilde{D} \nabla c) - \nabla \cdot (c\vec{v}) + P(c) - T(c,t) \quad (49)$$

where $T(c,t)$ takes into account terms that describe the application of therapy or a combination of therapies and exhibits an explicit time dependence according to the therapeutic protocol. The tumor growth model is considered in a general way including heterogeneous and anisotropic diffusion and advection terms combined with a proliferative term $P(c)$ which can be either exponential, logistic or Gompertz.

### a) Tumor resection therapy.

Tumor resection therapy was simulated as soon as the reaction diffusion model for glioma growth was proposed [4]. To account for this kind of therapy within the context of the reaction diffusion equation, tumor growth evolution is given by equation (49) and separated in two stages: Before tumor resection, which depends on tumor initial localization as an initial condition, and after tumor resection, assuming that the differential equation is subjected to the following condition [62],

$$c(\vec{r}, t_R) = F(\vec{r}) \qquad (50)$$

where $F(\vec{r})$ is the remnant distribution of tumor cells once the tumor is resected using the detectable tumor cell concentration threshold imposed by contrast enhanced CT or MRI, and $t_R$ is the resection time measured from the time the tumor start to grow in the simulation. Frequently, tumor resection therapy extracts a spherical tumor volume with an effective radius defined as,

$$r_R = \left(\frac{3}{4\pi} V_R\right)^{1/3} \qquad (51)$$

where $V_R$ is the tumor volume associated to those regions of the tumor cell concentration function that fulfill the condition $c(\vec{r}) \geq c_R$ with $c_R$, the threshold concentration used to assess tumor activity, i.e. defined by contrast enhanced CT or MRI. Results are shown in Figure 11 where two aspects have to be remarked: tumor recurrence is always expected since the margins defined by the detectable tumor cell concentration (typically defined by contrast enhanced MRI) do not include regions where disease is present (Figure 11a through Figure 11d, and that if tumor resection radius is increased by a 25 % (Figure 11 right side) a moderate increase in survival time, just a few weeks, is attained for a survival probability of 50 % compared to actual patient data that undergoes gross tumor resection [63].

### b) Chemotherapy.

Swanson et al. [64] proposed a method to quantify the efficacy of chemotherapy under the assumption of homogeneous and heterogeneous drug delivery. For the homogeneous case the starting equation is,

$$\frac{\partial c}{\partial t} = \nabla \cdot (\widetilde{D}\nabla c) + \rho c - G(t)c \qquad (52)$$

where $G(t)$ is a time dependent function defined by,

$$G(t) = \begin{cases} k, & t \in \{T_{on}\} \\ 0, & t \in \{T_{off}\} \end{cases} \qquad (53)$$

where $k$ is the strength of the action of the therapy drugs upon tumor cells and the sets $\{T_{on}\}$ and $\{T_{off}\}$ represent the collection of time intervals for which the drugs are administered (on) or not (off), respectively. Typically, chemotherapy is applied in treatment cycles, each one consists of the administration of drugs for a period of time $t_{chem}$ followed by a waiting period $t_{wait}$, meaning that the time dependence of $G(t)$ is an alternating function, periodic in most of the protocols used in chemotherapy, during the entire treatment time. The relevant parameter to determine the efficacy of the therapy is the ratio $\beta = k/\rho$ [64], which is a dimensionless parameter that compares the death rate caused by the drug to the proliferation rate, i.e. if $\beta > 1$, therapy effectively provides tumor control, see Figure 12a. In the case of heterogeneous drug delivery, equation (52) holds but $G(t)$ now becomes tissue dependent, i.e. gray and white matter, and is defined as,

$$G(\vec{r}, t) = \begin{cases} \begin{cases} k_W, & \vec{r} \in white\ matter \\ k_G = \alpha k_W, & \vec{r} \in gray\ matter \end{cases}, & t \in \{T_{on}\} \\ \qquad\qquad 0, & t \in \{T_{off}\} \end{cases}$$

(54)

where $\alpha$ is the ratio of capillary densities between gray and white matter, Figure 12b, c and d.. A more general model to account for drug delivery heterogeneity were proposed by Bratus et al. [65]. The authors assumed a coupled set of nonlinear reaction diffusion equations for the tumor cell concentration $c$ and the drug concentration, $h$,

$$\frac{\partial c}{\partial t} = \nabla \cdot (Dc^{2\alpha}\nabla c) + \rho c(1 - \beta lnc) - G(h)c \qquad (55)$$

$$\frac{\partial h}{\partial t} = \nabla \cdot (d_h h^{2\alpha_h}\nabla h) - \gamma_h h + u(t) \qquad (56)$$

where a Gompertz model was assumed for the tumor cell proliferation term in equation (55), $d_h$ and $\gamma_h$ are the drug diffusion coefficient and the drug elimination rate, respectively, $u(t)$, which is an explicit time dependent function, is the quantity of chemotherapeutic agent administered to the patient during a chemotherapy cycle, i.e. $u(t)$ follows the time protocol established in the chemotherapy treatment, and $G(h)$ is defined as,

$$G(h) = \frac{\lambda h}{1+h} \qquad (57)$$

For $h \ll 1$, equation (57) gives $G(h) \sim \lambda h$, and a linear dependence on the chemotherapeutic agent concentration is obtained. At higher concentrations chemotherapeutic activity departs from linearity and reaches its maximum strength or saturation level, $G \sim \lambda$.

### c) Radiotherapy.

Radiation therapy is usually modelled by the linear-quadratic or LQ radiobiological model [76,77], which assumes that the survival fraction of cells subjected to a total dose $D$ is given by,

$$S(D) = exp(-\alpha D - \beta D^2) \qquad (58)$$

where $\alpha$ and $\beta$ are parameters that characterize the response of tissue to ionizing radiation, which is determined by the damage that can be imparted to the DNA structure by the radiation. The damage depends on the Linear Energy Transfer or LET [76,77] of the ionizing particle (charged particles, neutrons or photons) and on which phase of the cell cycle, G2 or mitosis is the cell present. In order to avoid toxic effects of the radiation to normal tissue, radiotherapy is usually administered in fractions, and a biologically effective dose can be defined [77],

$$BED = nd\left(1 + \frac{d}{\alpha/\beta}\right) \quad (59)$$

with $n$, the number of fractions and $d$, the dose applied per fraction. In this case, the survival probability after the fractionated treatment is,

$$S(\alpha, \beta, d(t)) = exp(-\alpha BED) \quad (60)$$

Different schemes can be used for radiotherapy fractionation, $d(t)$, but typically, a fractionation scheme comprises 6 weeks of treatment, 5 days a week with weekend interruptions. Also, radiotherapy is a localized therapy, so the fraction dose is position dependent and it has to be taken into account during therapy simulation. Rockne et al. [67,69], Holdsworth et al. [70], Corwin et al. [72], Elazab et al. [74] and Kim et al. [75] proposed a model based on the following equation,

$$\frac{\partial c}{\partial t} = \nabla \cdot \left(\widetilde{D}\nabla c\right) + \rho c(1-c) - R(d(t))c(1-c) \quad (61)$$

where it has been emphasized the explicit time dependence of the fractionation scheme, $d(t)$, and $R(d(t))$ is defined as,

$$R(d(t)) = 1 - s(\alpha, \beta, d(t)) \quad (62)$$

and $s(\alpha, \beta, d(t))$ is the survival probability after one fraction dose,

$$s(\alpha, \beta, d) = exp\left(-\alpha d\left(1 + \frac{d}{\alpha/\beta}\right)\right) \quad (63)$$

There are two problems in relation to equation (61). The first one is related to the tumor cell concentration dependence of the radiation therapy term, i.e., a logistic one, which certainly assumes some negative "proliferation rate" given by equation (62), and for $c > 1$, introduces a positive contribution to tumor cell concentration, what is nonsense, so the radiation therapy term must be exponential with a negative cell death rate, $-R$. Nevertheless, equation (61) will reproduce qualitatively the typical behavior of the tumor size evolution, as shown in Figure 13. Additionally, equation (62) represents the cell death fraction due to radiation therapy and being a probability, it is a dimensionless quantity, so in order to keep all the terms in equation (61) with the same units, Rockne et al. [78] proposed that since fractionated radiotherapy is applied on a daily basis, equation (62) represents the probability of cell death during a time interval of 1 day and can be considered as a probability rate. Although this argument seems to be valid, Borasi et al. [73] developed a straightforward method to evaluate the cell death rate due to ionizing radiation, starting from the linear-quadratic model. If the radiation is applied at a constant dose rate, $\dot{d}$, during a time $\tau$, then according to the L-Q model, the cell concentration is,

$$c(\tau) = c(0) exp(-\alpha d(\tau) - \beta d(\tau)^2) =$$

$$c(0) exp(-\alpha \dot{d} \tau - \beta \dot{d}^2 \tau^2) \qquad (64)$$

where $d(\tau)$ is the administered dose at time $\tau$, then,

$$\frac{1}{c(\tau)} \frac{\partial c(\tau)}{\partial \tau} = -(\alpha \dot{d} + 2\beta \dot{d}^2 \tau) \qquad (65)$$

Taking into account that the irradiation time, $\Delta \tau$, is very small (of the order of minutes) compared to the time interval between fractions (one day), equation (65) can be replaced by its mean value over the irradiation time interval, $\Delta \tau$, so equation (61) must be replaced by [73],

$$\frac{\partial c}{\partial t} = \nabla \cdot (\widetilde{D} \nabla c) + \rho c(1 - c) - \dot{d}(\alpha + \beta d)c$$

(66)

Equation (66) is unit consistent and has the correct dependence of the therapy term on tumor cell concentration. Nevertheless, if the therapy term is considered alone, equation (66) does not recover the survival probability of the L-Q model, equation (64), so it must be replaced for one that includes the complete therapy term as given by equation (65),

$$\frac{\partial c}{\partial t} = \nabla \cdot (\widetilde{D}\nabla c) + \rho c(1-c) - \dot{d}(\alpha + 2\beta d(t))c$$

(67)

Due to the fact that $\widetilde{D}$ and $\rho$ are very small quantities, the temporal evolution of the tumor cell concentration imposed by the reaction diffusion terms during the time interval of radiation therapy can be considered as negligible, i.e. a suitable time interval for numerical integration of equation (67) is 1 day. Integration of equation (67) in a time interval very small compared to 1 day yields the correct survival probability of the L-Q model. This result allows for decoupling proliferation invasion from radiotherapy terms in equation (67) so at the end of 1 day of time evolution under radiotherapy conditions, tumor cell concentration is,

$$c(1\ day) = c_{PI}(1\ day)S(\alpha,\beta,d) \qquad (68)$$

with $c_{PI}$, the tumor cell concentration obtained from the proliferation invasion model. This approach has been used by Rojas et al. [59] and Unkelbach et al. [79,80] for high grade gliomas and by Henares-Molina et al. [81] for low grade gliomas.

The other aspect that has to be considered simulating radiation therapy is that it is a localized therapy and tumor delineation is very important for the therapy to be effective. Using the reaction diffusion model for glioma growth, Konukoglu et al. [82] demonstrated that the tumor cell concentration falls approximately in an exponential way depending on the distance, $|\vec{r}|$, measured from the detectable tumor interface, i.e., defined by contrast enhanced MRI,

$$c(|\vec{r}|) \propto exp\left(-\frac{|\vec{r}|}{\lambda_T}\right) \qquad (69)$$

where $\lambda_T$ is defined as the infiltration length and depends on the particular tissue, grey or white matter. It is related to the model parameters by,

$$\lambda_T = \sqrt{\frac{D_T}{\rho}} \tag{70}$$

Clearly, the infiltration length for white matter is bigger that the infiltration length for grey matter, $\lambda_W \sim \sqrt{D_W/D_G}$. Equation (70) allows for an improved tumor delineation using the infiltration length or multiples of it to establish a safety margin for the prescribed dose to be applied [80]. The estimation of subthreshold tumor based on the PI-DTI growth model has been proposed by Hathout et al. [83] as a tool for tumor delineation in radiation therapy planning. Unkelbach et al. [79] developed a treatment planning method that determines the cumulative dose $D$, distributed in $n$ fractions, which minimizes the survival probability within the tumor lesion, expressed as,

$$S(\alpha, \beta, d) = exp\bigl(-\alpha D(1 + \kappa D)\bigr) \approx exp(-\bar{\alpha} D) \tag{71}$$

where $\kappa = \beta/\alpha n$, $\bar{\alpha} = \alpha(1 + \kappa D_P)$, the effective radiosensitivity, and $D_P$ is the average prescription dose over the tumor lesion. The optimization problem is formulated as follows [79]. The integral cell survival,

$$f(D, c) = \sum_{i \in V_L} c_i exp(-\bar{\alpha}\, D_i) \tag{72}$$

is minimized subject to the condition,

$$\frac{1}{N}\sum_{i \in V_L} D_i \leq D_P \tag{73}$$

The sums are performed over all the voxels $i$ within the tumor lesion volume, $V_L$. The optimal solution is obtained by looking at the stationary points of the Lagrange function [79],

$$L = \sum_{i \in V_L} c_i exp(-\bar{\alpha} D_i) + \mu\bigl(\sum_{i \in V_L} D_i - N D_P\bigr) \tag{74}$$

where $\mu$ is the Lagrange multiplier and is given by [79],

$$\frac{1}{\bar{\alpha}} ln\left(\frac{\bar{\alpha}}{\mu}\right) = D_P - \frac{1}{\bar{\alpha} N}\sum_{i \in V_L} ln(c_i) \tag{75}$$

Results imposing the dose restrictions, given by equation (73), and the infiltration length dependence, given by equation (69), are shown in Figure 14. This approach has recently used by Lê et al. to establish a personalized treatment planning [84]. These results are applicable to define an optimal treatment planning and can be modified depending on the treatment specifics, i.e., single fraction or multifractioned treatment, IMRT or VMAT, and it minimizes the survival probability of the tumor lesion. In order to maximize the patient's survival probability, it is necessary to simulate a collection of virtual patients, all of them prepared with parameters derived from the actual patient, and subject them to the selected therapy. The patient's survival probability is then obtained by a Kaplan-Meier analysis [85] over the simulated course of the therapy. Rojas et al. [59,86] developed a method to estimate the effect of therapy over the patient´s survival probability, using the multi compartmental tumor growth model described in [58,59] and assuming additional transition rules imposed by the therapy, as shown in Table 2. Radiotherapy was considered as in equation (68), with a survival probability $S(\alpha, \beta, d)$ according to Table 3, depending if the tissue is normoxic, hypoxic or hypoglycemic. Results are shown in Figure 15.

**Conclusions.**

It has been shown that the proposition of a reaction diffusion model, starting from the basic proliferative-invasive Murray-Swanson model, to more detailed models that include brain anisotropy, tumor heterogeneity and biomechanical deformations, contributed in a significant way to support a working model suitable for simulation of patient specific therapy. Contributions, particularly in the area of chemotherapy and radiation therapy are very promising and possibly in a very near future will be available as routine methods in treatment planning and therapy optimization.

**Tables.**

Table 1. Voxel classification, thresholds and transition rules [59]

| Class | Description | Threshold values | Transitions |
|---|---|---|---|
| 1 | Proliferative-invasive state | $D > 0, c < 0.9$ | $1 \to 2$ |
| 2 | Proliferative-invasive and hypoxic state (reversible) | $D > 0, c \geq 0.9$ | $2 \to 1^*, 2 \to 3$ |
| 3 | Hypoglycemic state (reversible) | $\langle f \rangle \leq 0.02$ | $3 \to 2^\dagger, 3 \to 4$ |
| 4 | Necrotic state (irreversible) | $\langle f \rangle \leq 0.006$ | - |

Reversibility occurs: *If tumor cell concentration drops to $c < 0.9$, or †if $\langle f \rangle > 0.02$, due to fluctuations in the growth model only.

Table 2. Comparison of transition rules [86]

| Class | Description | Without therapy | With therapy |
|---|---|---|---|
| 1 | Normoxic | $1 \to 2$ | $1 \to 2$ |
| 2 | Hypoxic | $2 \to 1^*, 2 \to 3$ | $(2 \to 1^*), 2 \to 3$ |
| 3 | Hypoglycemic | $3 \to 2^\dagger, 3 \to 4$ | $(3 \to 2^\dagger), (3 \to 4)$ |
| 4 | Necrotic | - | - |

Reversibility occurs: *If tumor cell concentration drops to $c < 0.9$, or †if $\langle f \rangle > 0.02$, due to fluctuations in the combined growth and therapy model. Transitions within parenthesis are enhanced by therapy.

Table 3. Survival probability per fraction dose, $S(\alpha, \beta, d)$ [86]

| Class | $S(\alpha, \beta, d)$ |
|---|---|
| 1 | 0.83 [87] |
| 2 | 0.90 |
| 3 | 0.99 |

**Figure captions.**

Figure 1. Axial views of a glioma growth simulation using the proliferation invasion model or PI model at the time of diagnosis, left, and death, right, for an initiation site in frontal lobe, indicated by an asterisk. Color is associated to tumor cell concentration: red stands for high concentration and blue for low concentration. Black contour denotes the detection threshold by enhanced CT. Simulation elapsed time for diagnosis is 158 days and for death, 256 days. Adapted from [13,14].

Figure 2. Simulations using the PI-DT model performed on actual patient DT fractional anisotropy images. The anisotropy enhancement parameter $\kappa$, is zero for the isotropic case, shown on top. Adapted from [28].

Figure 3. Dependence of the Jaccard index with the tumor evolution time. The maximum in the graphic indicates when there is a match between the simulated and real tumors, allowing for assessment of tumor time and possibly tumor grading. Simulations, shown on the right, are represented over fractional anisotropy maps. The outline of the simulated tumor is represented in white while the detected tumor outline is represented in black. Adapted from [30].

Figure 4. Effect of biomechanical deformations, (a) Glioma growth simulation including diffusion tensor anisotropy and biomechanical deformations. (b) Brain tissue displacements induced by the tumor mas effect. In both graphics color indicates, red, highest tumor cell concentration or displacement and blue, for the lowest. Adapted from [19,20,31].

Figure 5. Tumor growth simulations on atlas space (right) registered on T1 reference images (left). Adapted from [39].

Figure 6. Results of simulations using equations (27-33). (a) Concentration profiles for different glioma grades. (b) Survival time as a function of $D$ and $\rho$. Adapted from [43].

Figure 7. Multi-compartmental model proposed by reference [51]. Compartments are C, proliferative or normoxic, H, hypoxic, Q, hypoglycemic and N, necrotic.

Figure 8. Search sphere used to determine nutrients concentration threshold for latency and necrosis states. Adapted from [59].

Figure 9. Seven years' time evolution for a virtual high grade glioma. On the left, distribution of tumor classes, according to Table 1, represented in color: red, class 2, yellow, class 3 and black, class 4. Right, time evolution of the number of voxels for each class. Adapted from [59].

Figure 10. Comparison of simulated GBM tumor progression models for a time period over 30 days with actual tumor images. (a) Observed tumor images, (b) simulation assuming anisotropic diffusion, without necrosis and (c) simulation assuming anisotropic diffusion with necrosis. Color indicates tumor cell concentration being red to denote the highest and violet for the lowest. Adapted from [61].

Figure 11. Simulations on tumor resection therapy. Left, (a) tumor concentration distribution defined by the condition $c(\vec{r}) \geq c_R$ at the resection time $t_R$, (b) remnant tumor concentration distribution, $F(\vec{r})$, immediately after resection (c) detectable tumor concentration distribution, $c(\vec{r}) \geq c_R$, 160 days after resection and (d) complete tumor concentration distribution function. Right, survival curves after resection: Actual patients subjected to gross total resection (GTR, asterisks), simulated patients subjected biopsy and no resection (BX/STR squares), simulated patients total resection of a volume defined by contrast enhanced MRI, with radius $r_{T_1}$ (GTR, circles) and simulated patients with total resection for $1.25 r_{T_1}$. Adapted from [62,63].

Figure 12. Efficacy of chemotherapy. (a) Size of the detectable tumor region as a function of time for different values of the parameter $\beta$, (b) A $T_2$-weighted equivalent map at the time of initiation of chemotherapy, (c) six weeks after the end of the chemotherapy cycles and (d) 30 weeks after (b). Notice the recurrence of disease. Adapted from [64].

Figure 13. Typical behavior of the time evolution of tumor size during radiation therapy. Blue line indicates tumor size evolution without therapy and red line with therapy. Arrows indicate (a) beginning and (b) end of the therapy, respectively.

Figure 14. Dose distributions that minimize tumor cell survival probability. Colors represent dose in Gy. Adapted from [79,80].

Figure 15. Simulations of radiotherapy for high grade gliomas. (a) Collection of untreated virtual patients with initiation point randomly selected in the right frontal lobe. (b) Tumor volume evolution with time. Different curves correspond to the application of therapy at different virtual years, characterized by a drop in the tumor volume. Vertical dashed line indicates the time when the untreated tumor reaches a volume that causes virtual patient death. (c) Survival probability obtained by a Kaplan-Meier analysis; thin continuous and dashed lines correspond to untreated tumors, thick continuous line corresponds to treatment.

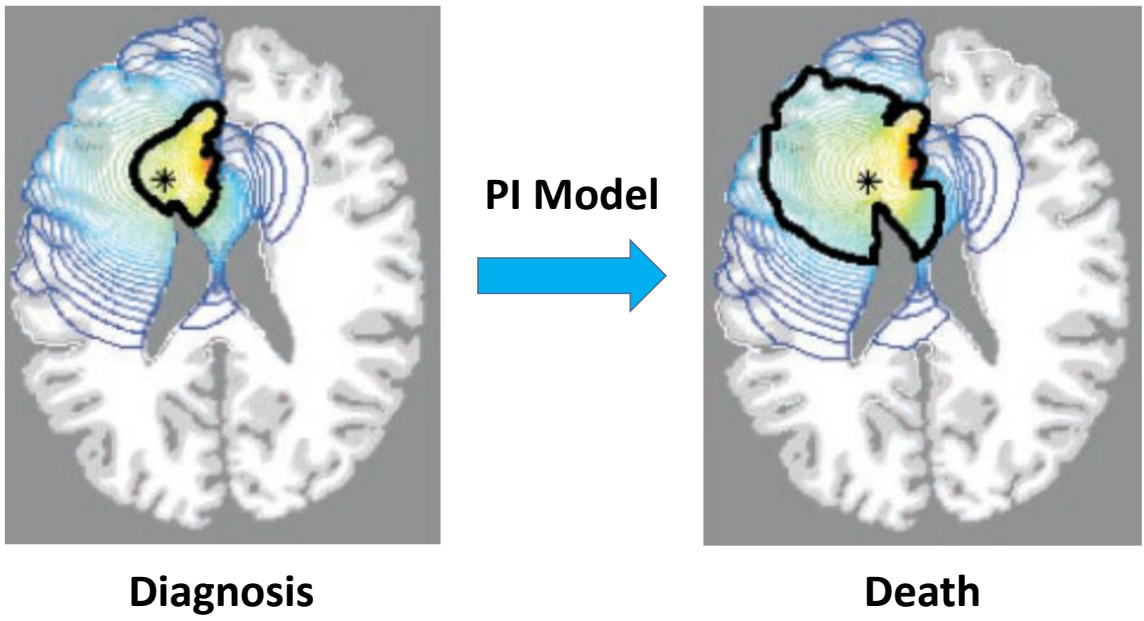

**Diagnosis**     **PI Model**     **Death**

Figure 1.

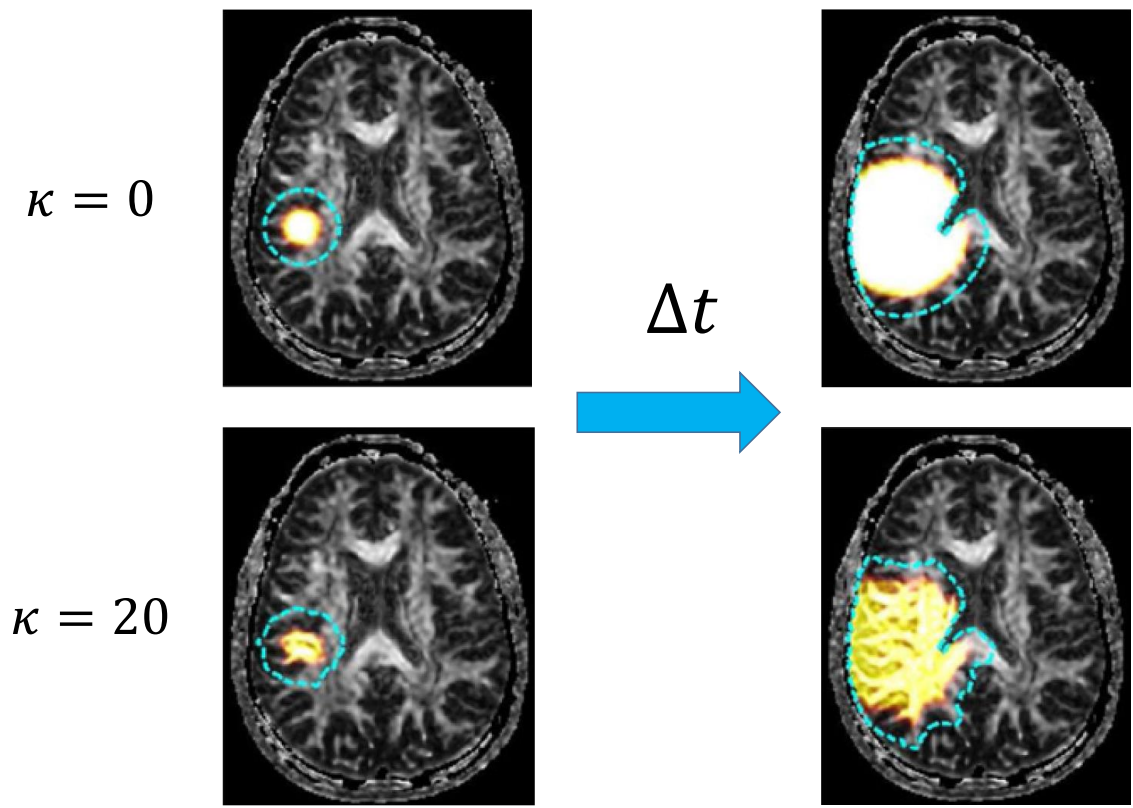

$\kappa = 0$

$\kappa = 20$

$\Delta t$

Figure 2.

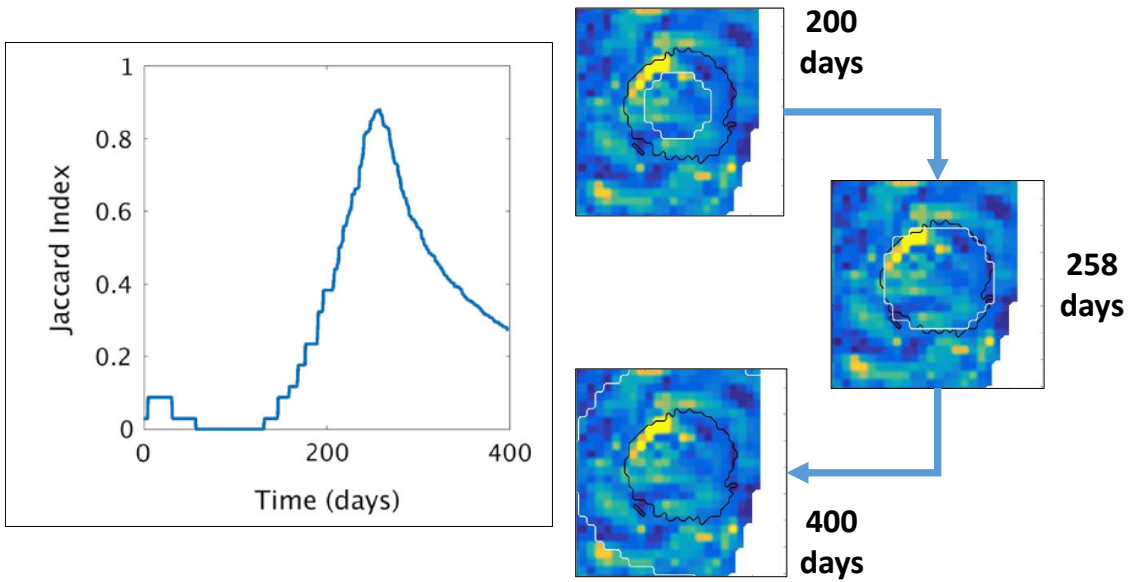

Figure 3.

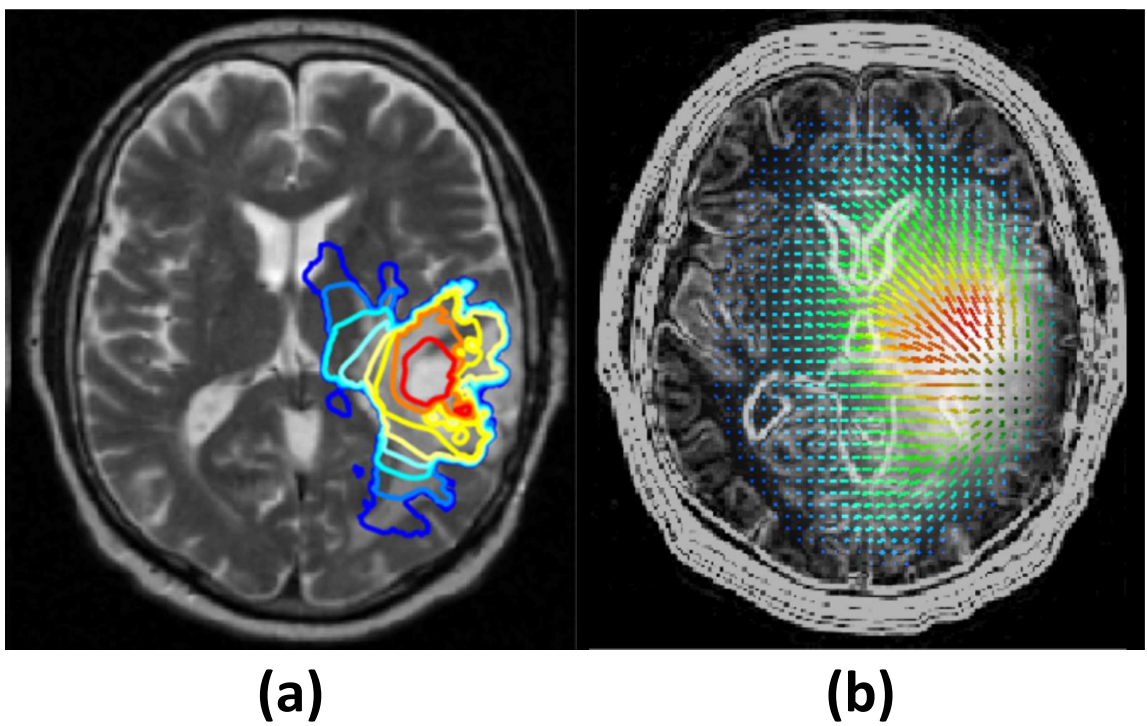

Figure 4.

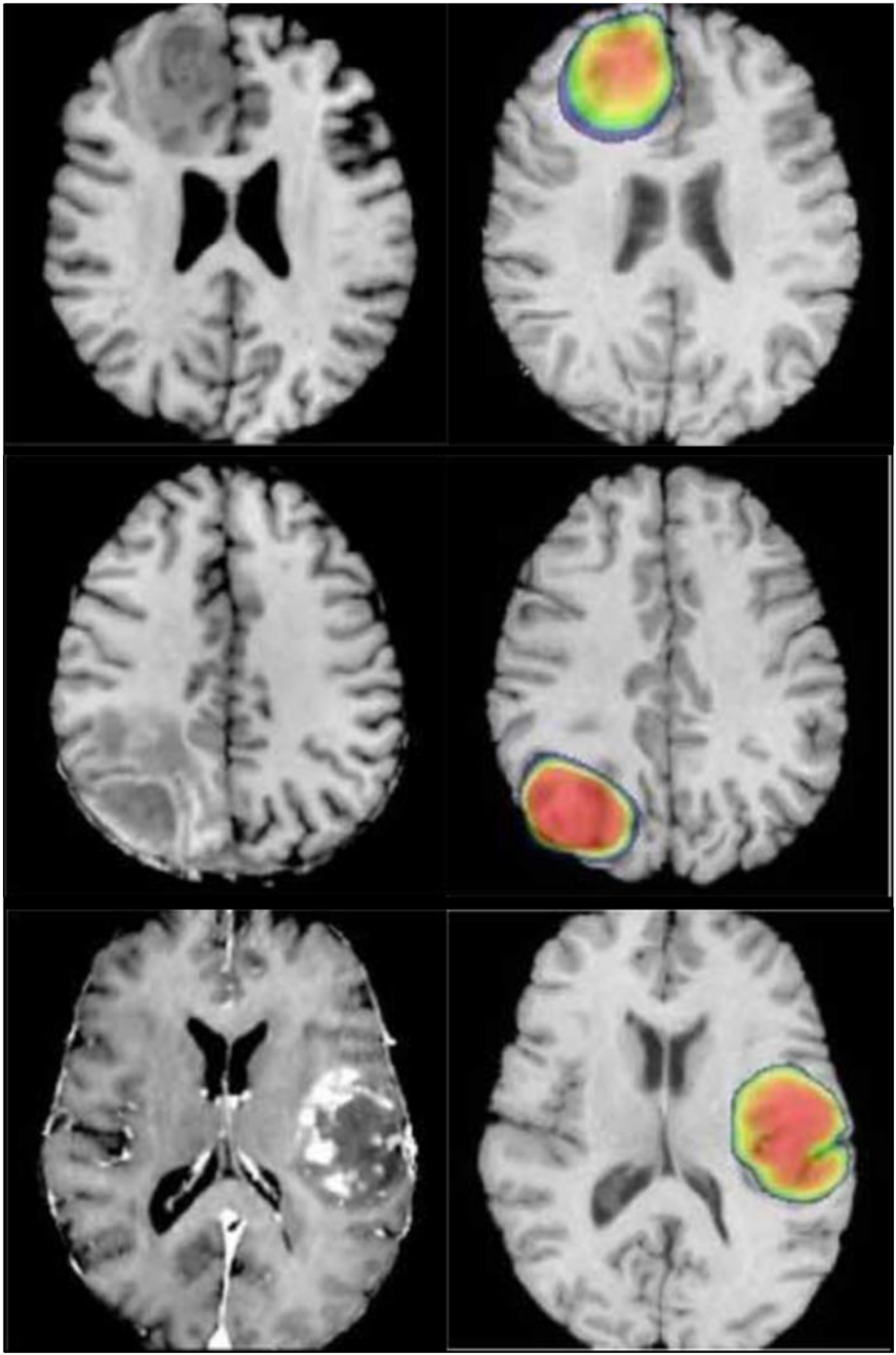

Figure 5.

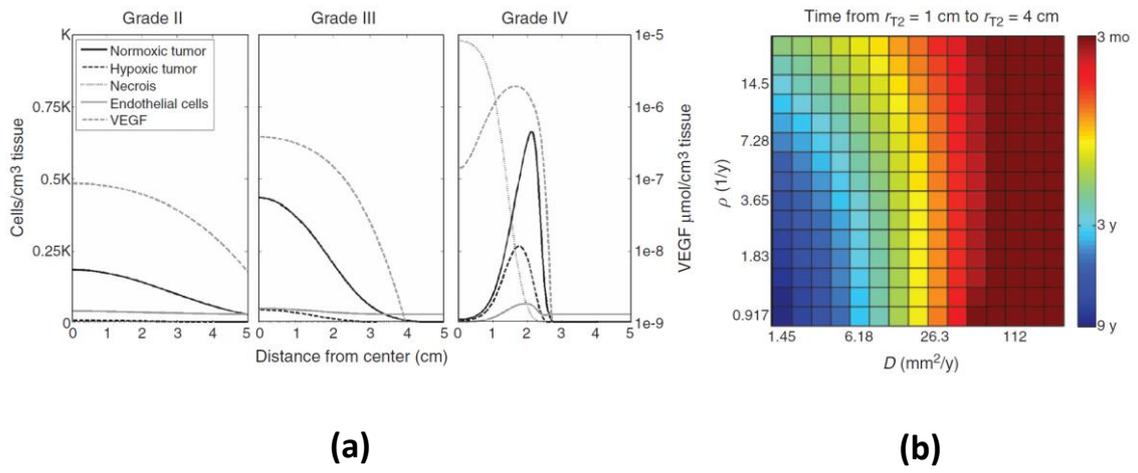

Figure 6.

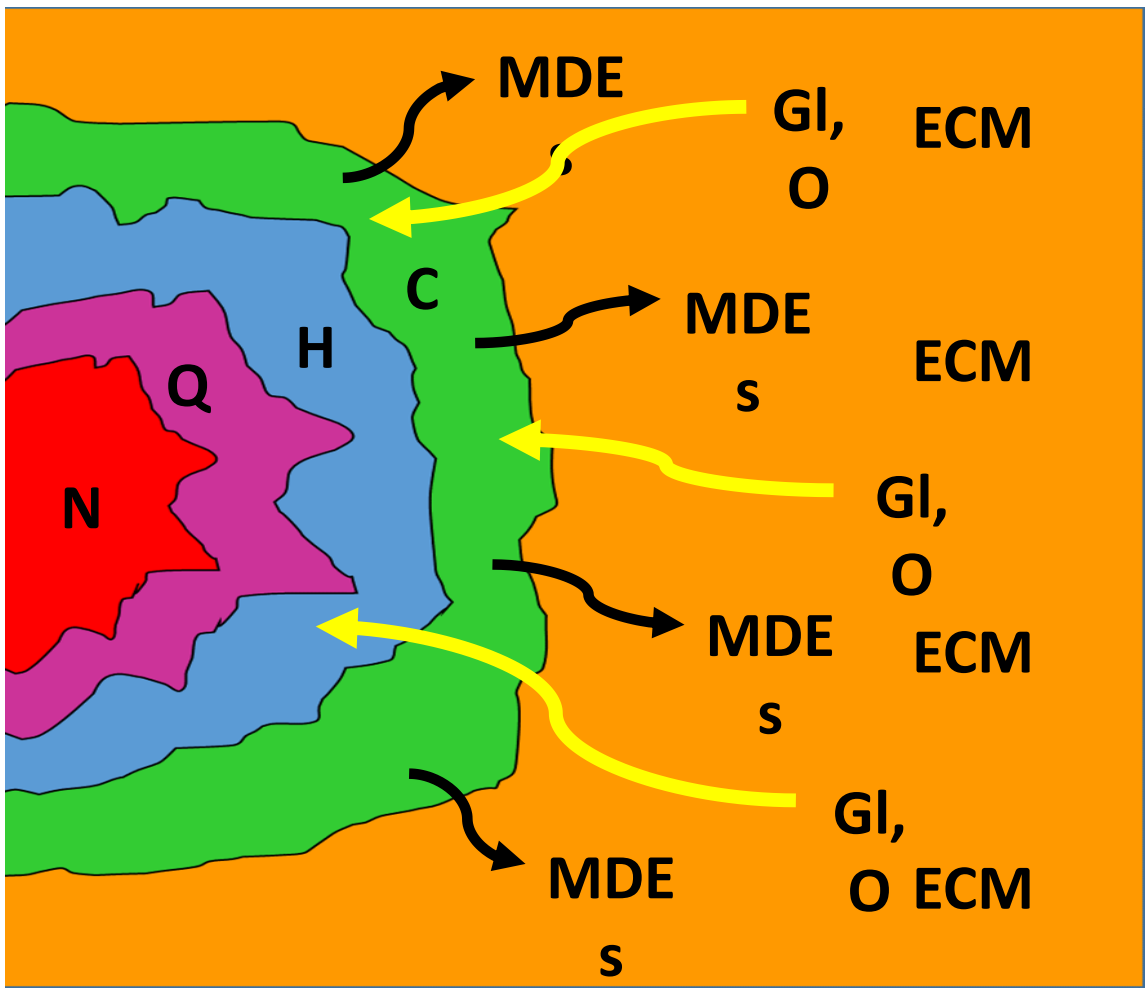

Figure 7.

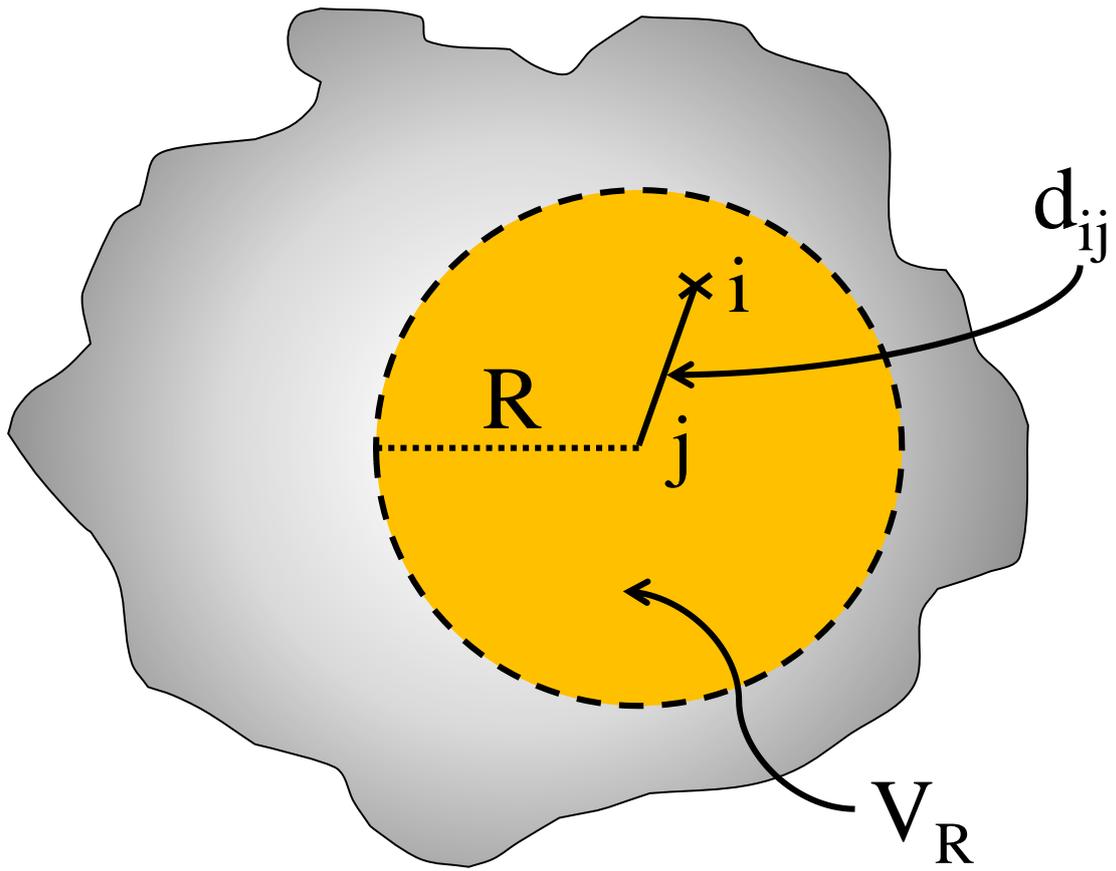

Figure 8.

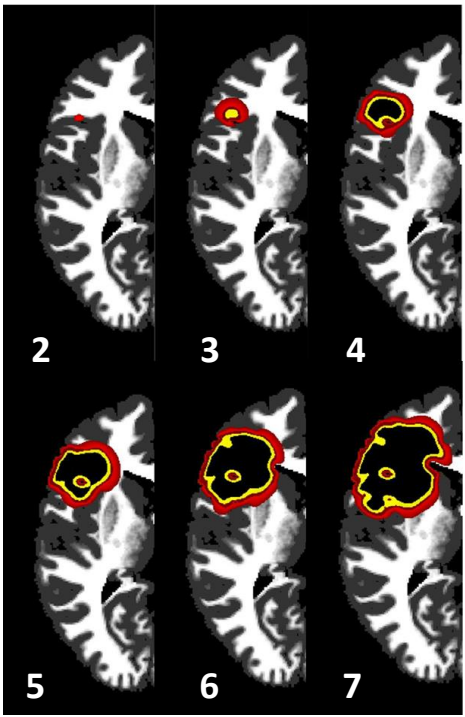
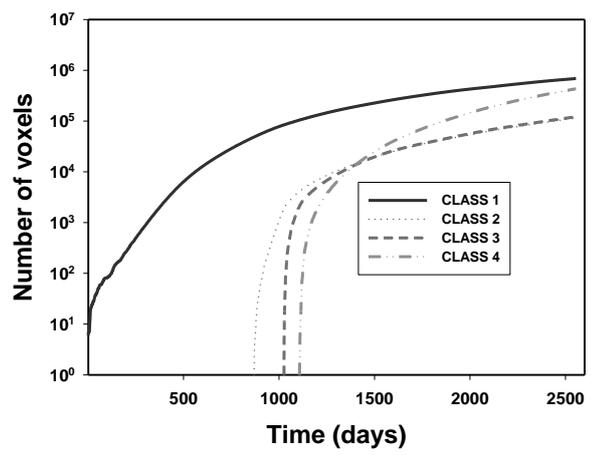

Figure 9.

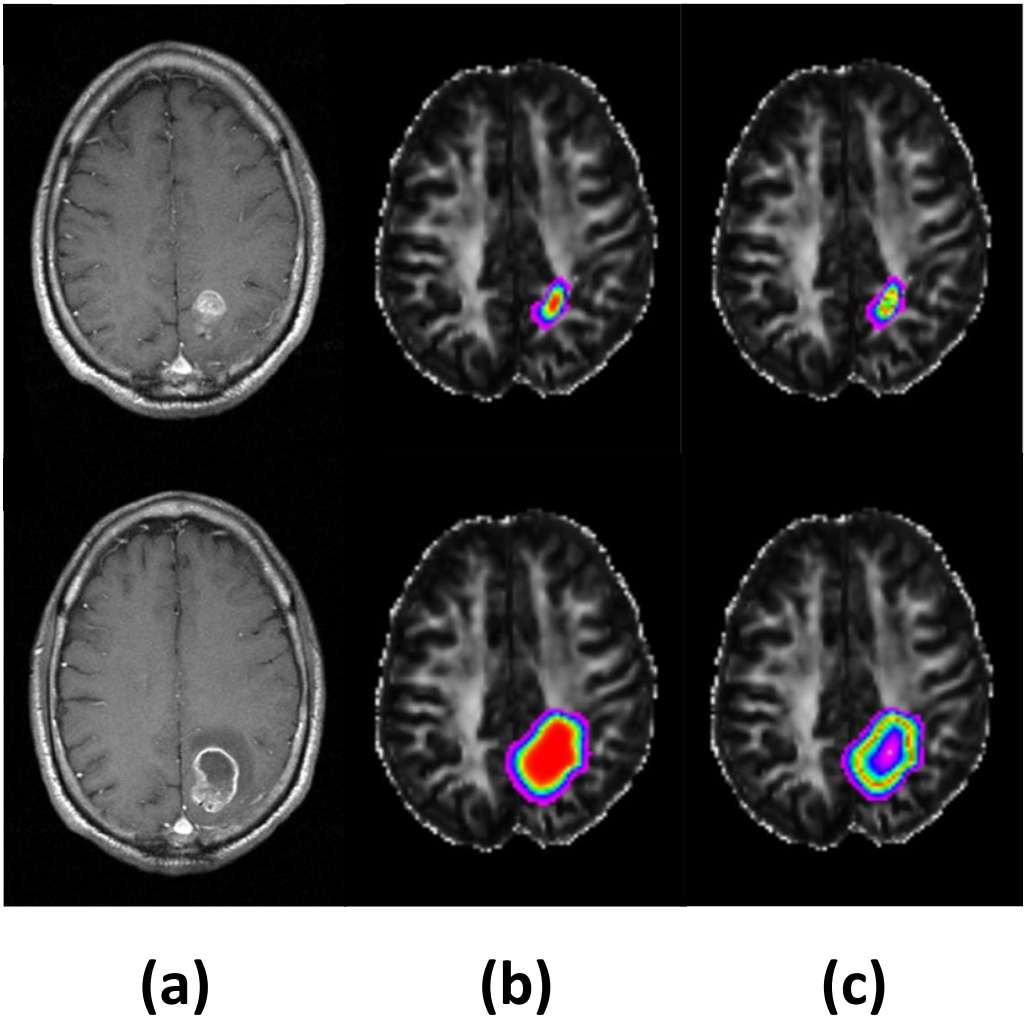

(a)          (b)          (c)

Figure 10.

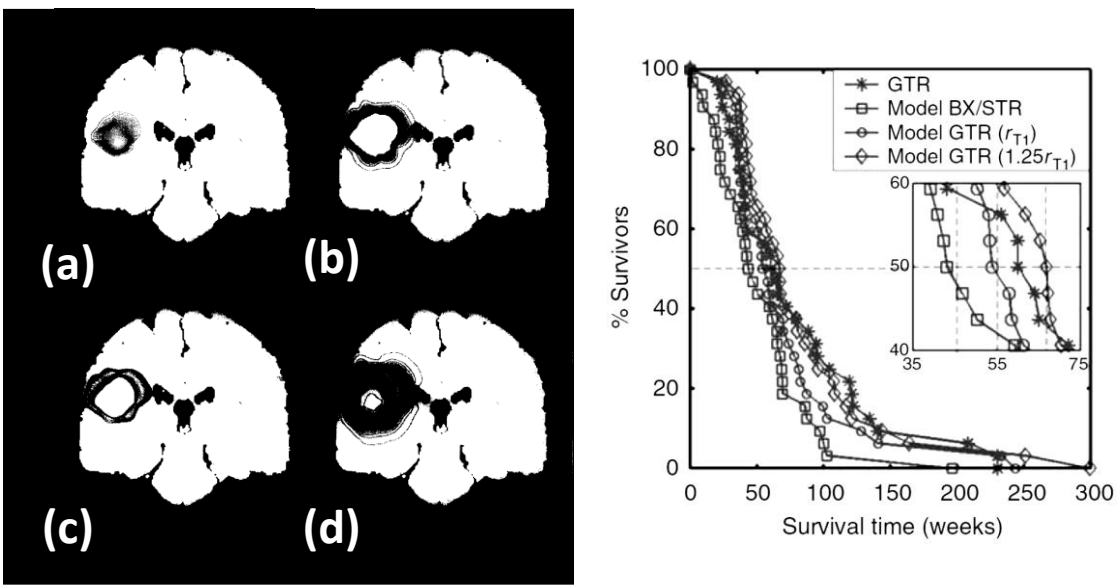

Figure 11.

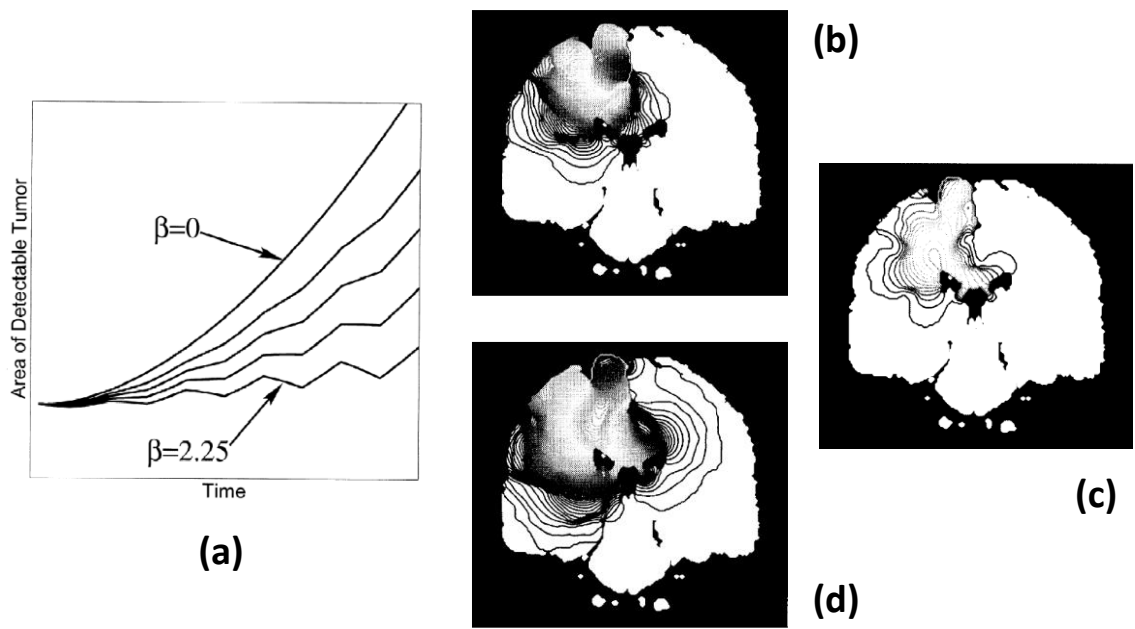

Figure 12.

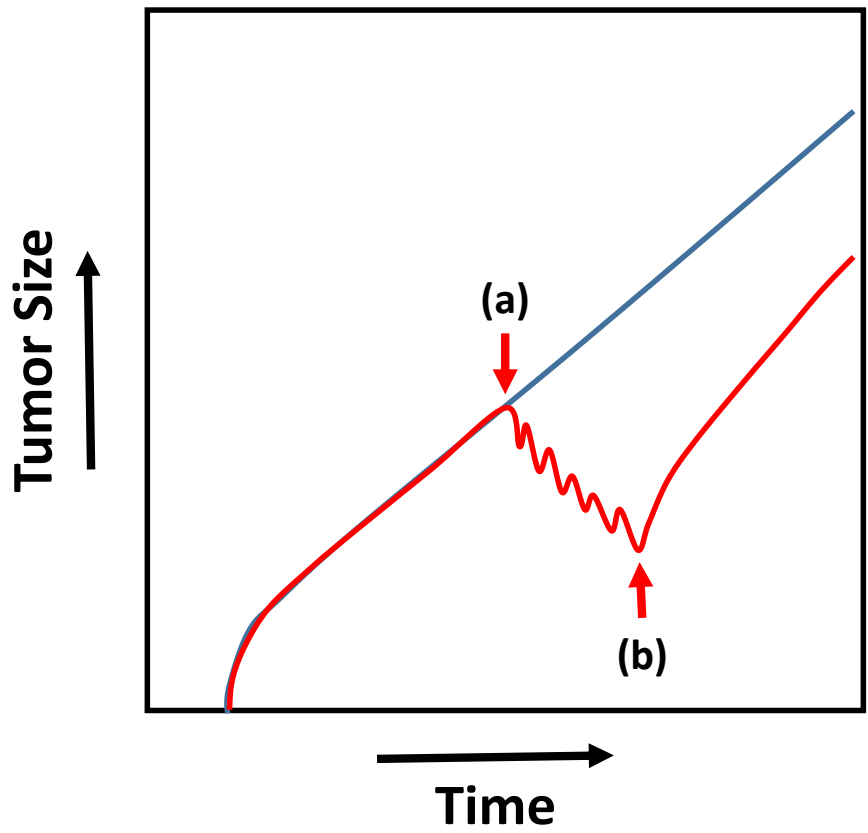

Figure 13.

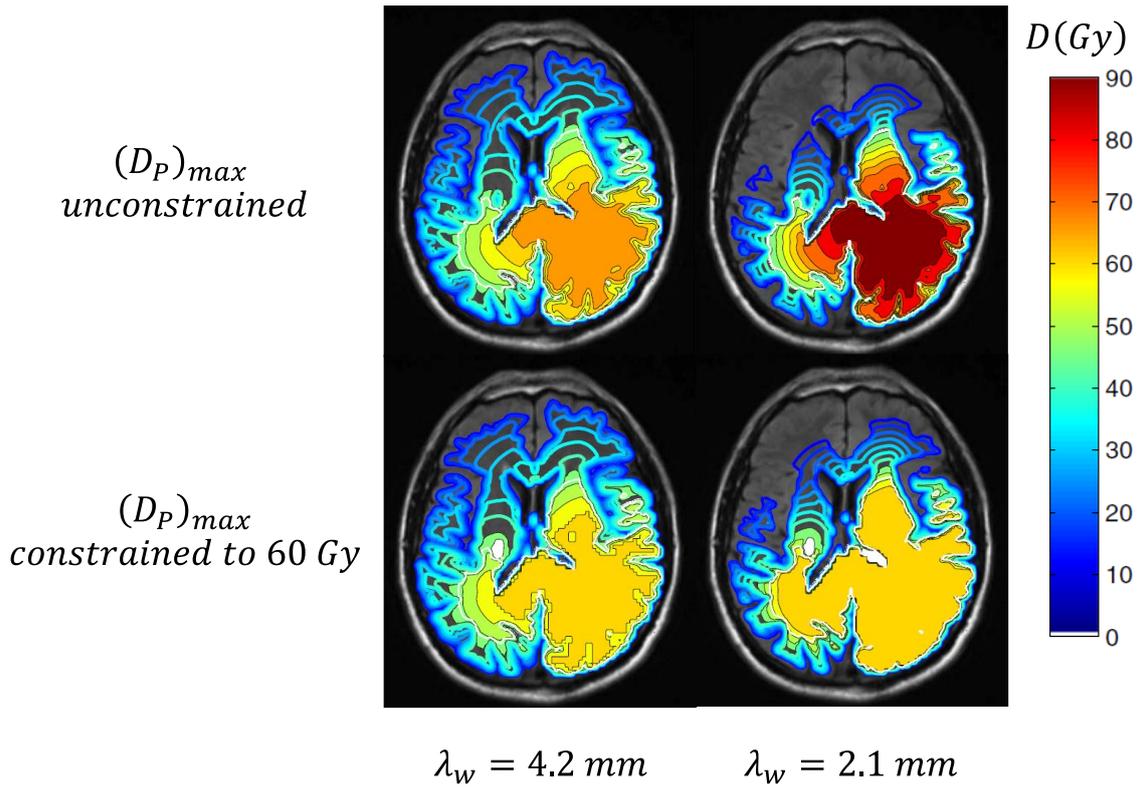

$(D_P)_{max}$
unconstrained

$(D_P)_{max}$
constrained to 60 Gy

$\lambda_w = 4.2\ mm$      $\lambda_w = 2.1\ mm$

Figure 14.

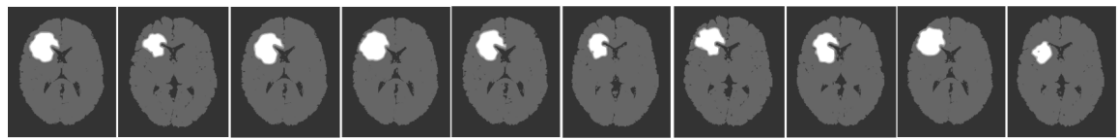

(a)

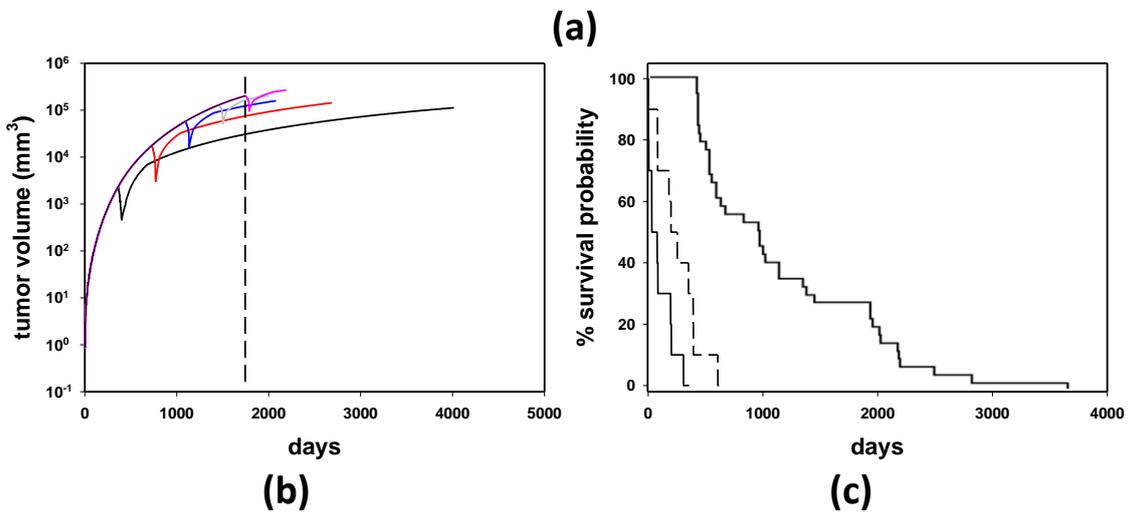

(b)         (c)

Figure 15.